\documentclass[10pt,conference]{IEEEtran}
\IEEEoverridecommandlockouts

\usepackage{cite}
\usepackage{amsmath,amssymb,amsfonts}
\usepackage{algorithmic}
\usepackage{graphicx}
\usepackage{textcomp}

\usepackage{booktabs} % fancy table
\usepackage{caption}
\usepackage{braket} % braket notation 
\usepackage{listings} % code listings
\usepackage{csquotes}

\usepackage[hyphens]{url}
\usepackage{hyperref}
\usepackage[hyphenbreaks]{breakurl}

\usepackage[svgnames]{xcolor}
  \definecolor{diffstart}{named}{Grey}
  \definecolor{diffincl}{named}{Green}
  \definecolor{diffrem}{named}{OrangeRed}

\definecolor{codegreen}{rgb}{0,0.6,0}
\definecolor{codegray}{rgb}{0.5,0.5,0.5}
\definecolor{codepurple}{rgb}{0.58,0,0.82}
\definecolor{backcolour}{rgb}{0.95,0.95,0.92}

\lstdefinelanguage{Ini}
{
    basicstyle=\ttfamily\small,
    columns=fullflexible,
    morecomment=[s][\color{Orchid}\bfseries]{[}{]},
    morecomment=[l]{\#},
    morecomment=[l]{;},
    commentstyle=\color{codegray}\ttfamily,
    morekeywords={},
    otherkeywords={=,:},
    keywordstyle={\color{codegreen}\bfseries}
}

\lstdefinelanguage{Python}{
    basicstyle=\ttfamily\footnotesize,
    morecomment=[f][\color{diffstart}]{@@},
    morecomment=[f][\color{diffincl}]{+\ },
    morecomment=[f][\color{diffrem}]{-\ },
    keywords={def,if,for,in, self, print},
}

\lstdefinestyle{mystyle}{
    language=Python,
    backgroundcolor=\color{backcolour},   
    commentstyle=\color{codegreen},
    keywordstyle=\color{magenta},
    numberstyle=\tiny\color{codegray},
    stringstyle=\color{codepurple},
    basicstyle=\ttfamily\footnotesize,
    breakatwhitespace=false,         
    breaklines=true,                 
    captionpos=b,                    
    keepspaces=true,                 
    numbers=left,                    
    numbersep=5pt,                  
    showspaces=false,                
    showstringspaces=false,
    showtabs=false,                  
    tabsize=2
}

\def\BibTeX{{\rm B\kern-.05em{\sc i\kern-.025em b}\kern-.08em
    T\kern-.1667em\lower.7ex\hbox{E}\kern-.125emX}}
    
\begin{document}
\bstctlcite{IEEEexample:BSTcontrol}

\title{Identifying Flakiness in Quantum Programs} 

\author{\IEEEauthorblockN{Lei Zhang}
\IEEEauthorblockA{\textit{Department of Information Systems} \\
\textit{University of Maryland, Baltimore County}\\
Baltimore, USA \\
leizhang@umbc.edu}
\and
\IEEEauthorblockN{Mahsa Radnejad}
\IEEEauthorblockA{\textit{Department of Computer Engineering} \\
\textit{Isfahan Branch, Islamic Azad University}\\
Isfahan, Iran \\
radnejad@khuisf.ac.ir}
\and
\IEEEauthorblockN{Andriy Miranskyy}
\IEEEauthorblockA{\textit{Department of Computer Science} \\
\textit{Toronto Metropolitan University}\\
Toronto, Canada \\
avm@torontomu.ca}
}

\maketitle

\begin{abstract}
In recent years, software engineers have explored ways to assist quantum software programmers. Our goal in this paper is to continue this exploration and see if quantum software programmers deal with some problems plaguing classical programs. Specifically, we examine whether intermittently failing tests, i.e., flaky tests, affect quantum software development.

To explore flakiness, we conduct a preliminary analysis of 14 quantum software repositories. Then, we identify flaky tests and categorize their causes and methods of fixing them.

We find flaky tests in 12 out of 14 quantum software repositories. In these 12 repositories, the lower boundary of the percentage of issues related to flaky tests ranges between 0.26\% and 1.85\% per repository. We identify 46 distinct flaky test reports with 8 groups of causes and 7 common solutions. Further, we notice that quantum programmers are not using some of the recent flaky test countermeasures developed by software engineers.

This work may interest practitioners, as it provides useful insight into the resolution of flaky tests in quantum programs. Researchers may also find the paper helpful as it offers quantitative data on flaky tests in quantum software and points to new research opportunities.
\end{abstract}

\section{Introduction}\label{sec:intro}

A test running on the same code sometimes produces different results, i.e., it shows ``passed'' sometimes and ``failed'' other times. Such tests are called flaky tests. Flaky tests can negatively impact developers by providing misleading signals. One can view flaky tests as bugs of testing that produce non-deterministic results. Such tests consume significant resources. At Google, in 2014, 73K out of 1.6M (4.56\%) of test failures were caused by flaky tests~\cite{luo2014empirical}; in 2017, 1.5\% of 4.2M tests were flaky~\cite{micco2017state, memon2017taming}.

A flaky test can be caused by two factors: either non-determinism in the source code or the test itself~\cite{luo2014empirical,eck2019understanding,parry2021survey,gruber2021empirical,dutta2020detecting,dutta2021flex}. Quantum programs are inherently non-deterministic. The randomness comes from a variety of sources. For example, it can be caused by the physical properties of quantum systems (e.g., quantum indeterminacy~\cite[Ch. 1]{marinescu2011classical}) or hardware issues (e.g., measurement errors or networking problems) or both (e.g., quantum decoherence~\cite[Ch. 7]{nielsen_chuang_2010}). In a simulation of a quantum computer (QC) on a classical computer (CC), pseudo-random number generators (PRNGs) are used to emulate these sources of randomness. Randomness from all these sources leads to a distribution of output values, which may result in flaky tests.

To the best of our knowledge, there exists no quantum software engineering research on flaky tests in quantum programs. Thus, we aim to do an initial analysis of flakiness and its root causes in quantum programs. To achieve this, we will seek answers to the following three \textbf{research questions}~(RQs). 

\textbf{RQ1}. How prevalent are flaky tests in quantum programs?

\textbf{RQ2}. What causes flakiness in quantum software?

\textbf{RQ3}. How do quantum programmers fix flaky tests?

The paper makes two major \textbf{contributions}. \textit{First}, we perform an initial study of flaky tests in quantum programs. Specifically, we investigate code and bug tracking repositories of 14 quantum software; 12 out of 14 software have at least one flaky test (46 unique flaky test reports in total). We estimate that at least 0.26\% to 1.85\% of issue reports in the 12 software are related to flaky tests. \textit{Second}, we identify and categorize eight groups of causes for flakiness in quantum software and seven common fixes. We find that the most common cause of flakiness is randomness, and the most common solution is to fix seeds for PRNGs. Moreover, quantum programmers do not use some recent countermeasures (e.g., ~\cite{dutta2020detecting,dutta2021flex}) developed by software engineers to deal with flaky tests. We publish the dataset that consists of quantum flaky tests, fixes, and categories at~\url{https://doi.org/10.5281/zenodo.7888639}.

The paper is organized as follows. Section~\ref{sec:empirical} presents our initial empirical study of flakiness in existing quantum programs. Section~\ref{sec:analysis} seeks answers to our three RQs by analyzing flaky test reports. Section~\ref{sec:threats} covers threats to validity. Section~\ref{sec:related} introduces related work of testing quantum programs and flaky tests in classical programs. Section~\ref{sec:conclusions} concludes the paper and outlines our long-term research objectives.

\section{Empirical Study: Data Gathering}\label{sec:empirical}

To answer our RQs, we perform an empirical study focusing on open-source quantum programs on GitHub because open-source projects have transparent development histories and bug reports. We follow three steps to collect flaky test reports from open-source quantum projects. 

First, we choose four popular quantum platforms (namely, IBM Qiskit~\cite{Qiskit:online}, Microsoft Quantum Development Kit~\cite{QandtheQ56:online}, TensorFlow Quantum~\cite{TensorFl88:online}, and NetKet~\cite{NetKetTh30:online}) and identify 14 repositories with active contributions and bug reports. Despite not being an exhaustive list of all quantum software, these platforms represent some of the most active and popular open-source quantum ecosystems.

Second, we searched closed Github issues associated with the 14 Github repositories for 10 keywords (namely, ``flaky'', ``flakiness'', ``flakey'', ``occasion'', ``occasional'',  ``intermit'', ``fragile'', ``non-deterministic'', ``nondeterministic'', and ``brittle''). We focus only on closed reports because they have been verified by developers. 

Finally, a minimum of two authors cross-examine these issues and associated pull requests and code commits, determine if they are related to flakiness, and establish the cause category. If there is a disagreement between the examiners, a joint review session is conducted to reach an agreement. In our case, the search for the 10 keywords returned 253 issue reports, 46 of which we manually labeled as flaky test reports (upon cross-examination). We then filter out two repositories without a verified flaky test (namely, \texttt{qiskit-finance} and \texttt{qiskit-optimization}) and end up with 12 repositories.

Based on our manual examination, the common case of a flaky test report consists of an issue report with a pull request that fixes the flakiness. However, there are three other cases: 1) multiple pull requests are related to a single report of flaky tests, e.g., a backport pull request, which we consider as one flaky test report; 2) a pull request that resolves flakiness without an issue report; and 3) a closed flakiness issue report without an associated pull request, e.g., a flaky test that is resolved without detailed conversations or corresponding commits. For simplicity, we call all four cases above ``flaky test reports''.   

\begin{table}[t]
    \centering
    \caption{Statistics of quantum software repositories with flaky tests. Three right-most columns are as follows: count of closed issue reports (Column $T$), count of closed flaky test reports (Column $F$), and percentage of reports related to flaky tests (Column $P$) computed as $F / T \times 100\%$.}
    \label{tab:repo}
    \begin{tabular}{@{}lllrrr@{}}    
    \toprule
    \emph{Platform} & \emph{Repository} & \emph{Language} & $T$ & $F$ & $P$ \\
    \midrule
    Qiskit & qiskit-terra    & Python & 2,810  & 25 & 0.89\% \\ 
    Qiskit & qiskit-aer    & Python & 558  & 3 & 0.54\% \\ 
    Qiskit & qiskit-nature & Python & 287 & 1 & 0.35\% \\ 
    Qiskit & qiskit-experiments & Python & 255 & 1 & 0.39\% \\
    Qiskit & qiskit-ibm-runtime & Python & 217 & 2 & 0.92\% \\
    Qiskit & qiskit-ibm-provider & Python & 184 & 2 & 1.09\% \\
    Qiskit & qiskit-machine-learning & Python & 378 & 1 & 0.26\% \\
    Microsoft & qdk-python & Python & 64 & 1 & 1.56\% \\
    Microsoft & QuantumLibraries & Q\# & 136 & 2 & 1.47\% \\
    Microsoft & Quantum & Q\# & 108 & 2 & 1.85\% \\
    TensorFlow & quantum & Python & 192 & 1 & 0.52\% \\
    NetKet & netket & Python & 295 & 5 & 1.69\% \\
    \midrule
    Total & & & 5,484 & 46 &   \\
    \bottomrule
    %\hline
    \end{tabular}
    %\vspace{1em}
\end{table}

\section{Analysis and Results}\label{sec:analysis}
In this section, we seek answers to our three RQs by analyzing the obtained flaky test reports.
\subsection{RQ1: How prevalent are flaky tests in quantum programs?}\label{sec:rq1}

Table~\ref{tab:repo} shows the statistics of the quantum program repositories and the flaky test reports that we detect. The data was last updated on January 12, 2023; thus, the statistics may change in the future. 

As shown in Table~\ref{tab:repo}, we detect 46 flaky test reports in the 12 repositories. Among all the repositories, the core Qiskit component \texttt{qiskit-terra} has the most flakiness reports (i.e., 25) because of the size of the project. The average percentage of flakiness varies from 0.26\% (\texttt{qiskit-machine-learning}) to 1.85\% (\texttt{Microsoft/Quantum}). Since our list of keywords is not exhaustive, the flakiness percentages represent a lower bound on the number of flaky test reports (see further discussion in Section~\ref{sec:threats}). In other words, there could be more flaky tests than what we have observed. 

Finally, comparing the frequency of flaky tests in QC and CC programs is difficult. As mentioned in Section~\ref{sec:intro}, Google reported percentages of test cases and test failures rather than percentages of flaky test reports; other authors typically report similar statistics (see~\cite{parry2021survey} for review). Therefore, a direct comparison is not possible. Our future plans include mapping flaky test reports to test cases and performing such a comparison. However, even without a direct comparison, we can definitively answer RQ1: flaky tests are present in significant quantities in quantum programs.

\subsection{RQ2: What causes flakiness in quantum software?}\label{sec:cause}

RQ2 is answered by manually analyzing all flaky test reports, categorizing them, and listing the summary in Table~\ref{tab:cause}. There are eight categories of causes with two special cases, i.e., ``others'' and ``unknown''. We detect one flakiness report containing issues and commits related simultaneously to randomness and floating point operations. Therefore, the total number of flaky test reports in Table~\ref{tab:cause} is 47 instead of 46 (the total number of flaky test reports in Table~\ref{tab:repo}).

What are the differences between our causes and those associated with CC software? Multiple analyses of flaky tests in CC software exist in the literature~\cite{luo2014empirical,eck2019understanding,parry2021survey,gruber2021empirical,dutta2020detecting}. For example, Luo et al.~\cite{luo2014empirical} identified 10 causes of flakiness in CC software: async wait, concurrency, test order dependency, resource leak, network, time, IO, randomness, floating point operations, and unordered collections. Most flaky tests in~\cite{luo2014empirical} are related to Java projects and distributed systems (e.g., Apache HBase and Hadoop).  We have five categories in common with theirs: concurrency, network, randomness, floating point operations, and unordered collections. 

The leading causes of flakiness in~\cite{luo2014empirical} are async waits and concurrency. Our top causes of flaky test reports are randomness (21\%) and software environments (15\%). Why is this discrepancy occurring?

Researchers have identified randomness as a non-major cause of flakiness in generic CC software~\cite{luo2014empirical, parry2021survey, gruber2021empirical}. A survey~\cite{parry2021survey} indicates that in 1\% to 5\% of cases (depending on the software), flaky tests are caused by randomness, while async wait and concurrency are consistently the leading causes.

Our findings are closer to those focusing on probabilistic programming and machine learning software, where 60\% of flaky tests may be caused by randomness~\cite{dutta2020detecting}. As PRNGs are heavily used in QC programming (see Section~\ref{sec:indeterminacy}), randomness contributes significantly to flaky test reports in quantum programs.

Let us examine the eight causes (as well as the two special cases) listed in Table~\ref{tab:cause} in more detail.

\begin{table*}[t]
\centering
\caption{Count of cause categories and fix patterns based on flaky test reports.}
\label{tab:cause}
\resizebox{\textwidth}{!}{\begin{tabular}{@{}l|rrrrrrrr|rr@{}}
\toprule
                     & \multicolumn{8}{c|}{Fix pattern}                                                                                                          &              &  \\ \cmidrule(lr){2-9}
Cause category       & Fix Seed & Alter Software Env. & Make Single Thread & Adjust Tolerance & Add Exception Handler & Synchronize & Use Keys for Order & Others & Grand Total & Percentage    \\ \midrule
%        & Fix  & Alter & Make & Adjust  & Add Exception &  & Use Keys &  & &     \\ 
% Cause category       &  Seed & Softw. Env. &  Single Thread &  Tolerance & Handler & Synchronize & for Order & Others & Grand Total & Percentage    \\ \midrule
Randomness           & 9        &                     &                    &                  &                       &             &                    & 1      & 10          & 21\%  \\
Software Env.        &          & 4                   &                    &                  &                       &             &                    & 3      & 7           & 15\%  \\
Multi-Threading      &          &                     & 2                  &                  &                       &             &                    & 4      & 6           & 13\%  \\
Floating Point Ops.  &          &                     &                    & 3                &                       &             &                    & 2      & 5           & 11\%  \\
Visualization        &          &                     &                    &                  &                       &             &                    & 3      & 3           & 6\%   \\
Unhandled Exception  &          &                     &                    &                  & 3                     &             &                    &        & 3           & 6\%   \\
Network              &          &                     &                    &                  &                       & 1           &                    &        & 1           & 2\%   \\
Unordered Collection &          &                     &                    &                  &                       &             & 1                  &        & 1           & 2\%   \\
Others               &          &                     &                    &                  &                       &             &                    & 7      & 7           & 15\%  \\
Unknown              &          &                     &                    &                  &                       &             &                    & 4      & 4           & 9\%   \\ \midrule
Grand Total          & 9        & 4                  & 2                  & 3                & 3                     & 1           & 1                  & 24     & 47          & 100\% \\
Percentage           & 19\%     & 9\%                & 4\%                & 6\%              & 6\%                   & 2\%         & 2\%                & 51\%   & 100\%       &      \\ \bottomrule
                     
\end{tabular}}
\end{table*}

\subsubsection{Randomness}\label{sec:indeterminacy}

As mentioned above, the use of PRNGs is the most common (21\%) cause of reports of flakiness in quantum programs. The PRNGs produce different outputs from run to run, which may result in a flaky test. Given that most of the test cases are executed on simulators of QCs and that the simulators rely heavily on PRNGs to simulate the non-deterministic nature of QCs, it is not surprising that this is the primary cause of flaky test reports.

PRNGs are used in quantum programs and associated test suites. Our first example (Listing~\ref{lst:randomness}) demonstrates a problem with the quantum program found in issue report \#3533 of \texttt{qiskit-terra}. The constructor of a two-qubit decomposer (details in~\cite{PhysRevA.63.062309}) has a non-deterministic component in Lines 6 and 7. This decomposer can generate random results when generating controlled versions of the same unitary gate. For example, Listing~\ref{lst:unitary-test} creates a \texttt{UnitaryGate} instance in Line 1, and the equality operation in Line 5 returns ``True'' or ``False'' at random. 

\begin{lstlisting}[language=Python, label={lst:randomness},
caption=An example of randomness in a quantum program that can lead to a flaky test.]
class TwoQubitWeylDecomposition:
    ...
    def __init__(self, unitary_matrix):
        ...
        for _ in range(100): 
            M2real = np.random.randn()*M2.real \
                + np.random.randn()*M2.imag
\end{lstlisting}

\begin{lstlisting}[language=Python, label={lst:unitary-test},
caption=A testing code used to reproduce the flaky test.]
uni = UnitaryGate([[1, 0, 0, 0],
                   [0, 1, 0, 0],
                   [0, 0, 1, 0],
                   [0, 0, 0, 1j]])
uni.control() == uni.control()
\end{lstlisting}

\begin{lstlisting}[language=Python, label={lst:seed}, caption=An example of a PRNG-related flaky test in the code of a test case.]
def test_append_circuit(self, num_qubits):
    ...
    first_circuit = random_circuit(num_qubits[0], depth)
    ...
    for num in num_qubits[1:]:
        circuit = random_circuit(num, depth)
\end{lstlisting}

In Listing~\ref{lst:seed}, the second example (based on issue report \#5217 in \texttt{qiskit-terra}) illustrates a test case issue. The logic of the test \texttt{test\_append\_circuit}, which checks if appending quantum circuits work properly, is correct. However, Lines 3 and 6 cause flakiness in the test because the function \texttt{random\_circuit} generates a random quantum circuit using a randomly selected seed (by default). Thus, the test fails occasionally due to randomness. We will discuss how to fix the flakiness in Listings \ref{lst:randomness} and \ref{lst:seed} in Section~\ref{sec:fixed-seed}.

\subsubsection{Software Environment}
This category includes flaky tests caused by specific software or library dependency issues. For example, pull request \#1369 in \texttt{netket} discusses a flaky test observed only in the GitHub Actions Python 3.10 environment. The pull request starts with the following comment. 

\begin{displayquote}
``No idea why, but on the GitHub runner python 3.10 this test keeps failing. I never reproduced locally so I think it's not a real failure. Simplyfying [\textit{sic}] the test to avoid.''
\end{displayquote}

A common way to fix this issue is to alter the software environment as will be shown in Section~\ref{sec:alter_softw_env}.

\subsubsection{Multi-Threading}
This category of flaky tests is caused by multi-threading issues, e.g., concurrency and overload. As an example, issue report \#5904 in \texttt{qiskit-terra} describes a flaky test caused by address collisions due to parallel builds. The code in this example can be seen in Listing~\ref{lst:multi-thread}, showing that \texttt{sphinx-build} generates documentation in parallel over $N$ processes, where $N$ is the number of CPUs (i.e., by the argument \texttt{-j auto}). While the parallelization improves the throughput, it causes an error when multiple jobs of \texttt{sphinx-build} compete against each other.  A fix pattern for this cause is given in Section~\ref{sec:make_single_thread}.

\begin{lstlisting}[language=Ini, label={lst:multi-thread},
caption=An example of a flaky test related to multi-threading.]
commands = sphinx-build -W -b html -j auto docs/ docs/_build/html {posargs}
\end{lstlisting}

\subsubsection{Floating Point Operations}\label{sec:precision}

We define a flaky test as floating-point-related if it occurs due to challenges such as round-off errors. For example, in \texttt{netket} pull request  \#1147 (Listing~\ref{lst:netkit_assert}), the hard-coded tolerance value of $10^{-5}$ (shown in Line 5) causes a test case to fail intermittently.

\begin{lstlisting}[language=Python, label={lst:netkit_assert},
  caption=An example of a flaky test related to floating point operations.]
def test_vmc_functions():
    ha, sx, ma, sampler, driver = _setup_vmc()
    driver.advance(500)
    assert driver.energy.mean == \ 
        approx(ma.expect(ha).mean, abs=1e-5)
\end{lstlisting}

To tackle this problem, developers modify the assertion to alter tolerance, round the actual value, or remove a flaky test case altogether.
We will provide details of how developers address this challenge in Section~\ref{sec:increase-tolerance}.

\subsubsection{Visualization}
This group of flaky tests is related to image generation. For example,  \texttt{qiskit-terra} has a test manager that schedules visual tests sequentially and allows these tests to communicate with one another. Issue \#3283 in \texttt{qiskit-terra} reports a flaky test in the visualizer of the test manager due to out-of-date reference indexes.

In this category, we are unable to identify a recurring fix pattern. There is a unique solution for each of the three reports (all associated with \texttt{qiskit-terra}).

\subsubsection{Unhandled Exception}\label{sec:exception}
This group of flaky tests is caused by the code that does not appropriately handle exceptions. For example, Listing~\ref{lst:unhandled} shows the stack trace in the issue report of \#398 in \texttt{Microsoft/QuantumLibraries}. In this case, the testing can occasionally fail whenever the ``number'' does not fall into the expected range. We will explore a fix pattern for this cause in Section~\ref{sec:add_exception_handle}.

\begin{lstlisting}[language=bash, label={lst:unhandled}, caption=An example of an unhandled exception.]
Unhandled exception.
Microsoft.Quantum.Simulation.Core.ExecutionFailException: "number" must be between 0 and 2^3 - 1, but was -1.
\end{lstlisting}

\subsubsection{Network}
A flaky test in this category occurs as a result of network-related issues, such as an unstable network or server. For example, issue \#584 of \texttt{qiskit-ibm-runtime} reports a flaky test due to timeouts and socket connection problems. As shown in Listing~\ref{lst:network}, \texttt{test\_websocket\_proxy} fails because jobs have been completed before \texttt{websocket} connection can be established. A method to fix this cause is given in Section~\ref{sec:synchronize}.

\begin{lstlisting}[language=bash, label={lst:network}, caption=An example of network-related flaky test.]
FAIL: test_websocket_proxy (test.integration.test_results.TestIntegrationResults) 
    (service=<QiskitRuntimeService>)
\end{lstlisting}

\subsubsection{Unordered Collection}
This category of flaky tests is Python-specific (although, hypothetically, it may appear in other languages). Dictionaries (hash maps or hash tables) are implemented as an unordered collection from Python 3.3 to 3.7~\cite{gruber2021empirical}. When tests have order dependencies, test outcomes can become non-deterministic, which leads to flakiness. For example, pull request \#8627 in \texttt{qiskit-terra} reveals a flaky test that uses insertion order to compare two dictionaries, hence non-deterministic results. We will explore a fix for this cause in Section~\ref{sec:use_keys_order}.

\subsubsection{Others}
Flaky test reports are included in this special category if there is only one observation of the cause and the cause has not been classified in previous studies (e.g.,~\cite{luo2014empirical, parry2021survey, gruber2021empirical}). 

For example, in Listing~\ref{lst:others} (issue report \#62 in \texttt{Microsoft/Quantum}), there is a space between the left double quotation mark and \texttt{Microsoft}, which accidentally injects a random value into the test and causes intermittent failures. 

\begin{lstlisting}[language=Python, label={lst:others}, caption=A typo causes a flaky test.]
fixedSeeds.Add(" Microsoft.Quantum.Tests.RobustPhaseEstimationTest", 2020776761);
\end{lstlisting}

Another example is keeping local outputs in Jupyter Notebook and preventing flaky tests from being executed in the continuous integration pipeline (pull request \#453 in \texttt{TensorFlow/quantum}).

A final example is given in pull request \#795 of \texttt{qiskit-aer}. It relates to the changing configuration of a QC (as engineers keep tweaking the computer's configuration). Engineers update the configuration files of the simulator of this QC whenever the configuration changes. However, the test case expected values were hard-coded, which caused intermittent test case failures. To resolve this issue, testers started dynamically reading configuration parameters from configuration files instead of hard-coding them.

\subsubsection{Unknown}
A flaky test cause is classified in this special case as unknown if there is insufficient information about its cause or fix, i.e., it is impossible to identify the cause of a flaky test without detailed conversations or corresponding commits. For example, a flaky test stack trace is provided in issue report \#185 of \texttt{qiskit-machine-learning}. The report is later closed because of insufficient information.

\subsection{RQ3: How do quantum programmers fix flaky tests?}\label{sec:fix}

Table~\ref{tab:cause} shows seven fix patterns for the flaky test reports that we found (based on the flakiness-related commits and pull requests). These patterns cover 49\% of fix reports. The rest of the fixes are one-off and are bundled into ``others'' fix pattern (51\%). The one cause category that has no fix pattern associated is ``visualization'' because there exists no recurring pattern and we only observe visualization-related flaky tests in \texttt{qiskit-terra}. 

\subsubsection{Fix Seed}\label{sec:fixed-seed}
It is easier to compare a variable with a constant when a seed used to initialize a PRNG is fixed. Thus, 9 out of 10 randomness-related flaky test reports (i.e., 19\% of all flaky test reports) are fixed by setting a random seed value to a constant (or in combination with a reduced convergence threshold as shown in pull request \#8820 of \texttt{qiskit-terra}). For example, to avoid flakiness in Listing~\ref{lst:seed}, Listing~\ref{lst:fix-seed} (pull request \#5599 in \texttt{qiskit-terra}) shows a solution that replaces the default non-constant random seed value with a fixed seed value. 

\begin{lstlisting}[language=Python, label={lst:fix-seed}, caption=An example of fixed seed for Listing~\ref{lst:seed}.]
- first_circuit = random_circuit(num_qubits[0], depth)
+ first_circuit = random_circuit(num_qubits[0], depth, seed=4200)
...
- circuit = random_circuit(num, depth)
+ circuit = random_circuit(num, depth, seed=4200)
\end{lstlisting}

While flakiness can be controlled by fixing seeds for the PRNGs, this approach can potentially make testing less effective because it limits possible execution paths that can expose real bugs~\cite{dutta2021flex}. Further, a change in the PRNG algorithm may make the test case flaky again.

We have not seen robust fixes to this issue that involve running the test case multiple times to compute the distribution\footnote{Although we saw a strategy of running the test three times and declaring success if the test passes at least once (pull request \#724 in \texttt{netket}).} of successful and failed executions, then performing distribution analysis to assess whether new changes to the code alter the distribution (see~\cite{dutta2021flex} for details). This could be due to higher computation and execution time requirements. Additionally, programmers may not want to build such a test framework from scratch and do not realize that test frameworks like this, e.g., FLEX~\cite{dutta2021flex}, already exist.

Finally, we find one flaky test that is fixed by replacing the PRNG with a deterministic formula (pull request \#3585 of \texttt{qiskit-terra}, as shown in Listing~\ref{lst:remove-rng}). 

\begin{lstlisting}[language=Python, label={lst:remove-rng}, caption=An example of removing the source of randomness in Listing~\ref{lst:randomness}.]
- for _ in range(100):
-   M2real = np.random.randn()*M2.real + np.random.randn()*M2.imag
+ for i in range(-50, 50):
+   M2real = (i/50)*M2.real + (i/25)*M2.imag
\end{lstlisting}

\subsubsection{Alter Software Environment}\label{sec:alter_softw_env}
Flaky tests can be removed by updating or changing the development environment. In the seven observed flaky test reports related to the software environment (see Table~\ref{tab:cause}), four of them (9\% of all the reports) are fixed by upgrading or changing the dependencies. Listing~\ref{lst:fix-env} demonstrates a fix by removing the dependency on a specific \texttt{numpy} version (more details can be found in pull request \#4656 of \texttt{qiskit-terra}). 

\begin{lstlisting}[language=Python, label={lst:fix-env}, caption=An example of fixing library dependency.]
- numpy!=1.19
\end{lstlisting} 

The remaining three reports are fixed by changing the configuration of the continuous integration pipeline (issue \#319 in \texttt{Microsoft/qdk-python}), simplifying the test case (pull request \#1369 in \texttt{netket}), or simply closing the report because of dependency changes (issue \#1466 in \texttt{qiskit-aer}).

\subsubsection{Make Single Thread}\label{sec:make_single_thread}
Multi-threading-related flaky tests can be resolved by limiting the number of threads to one. Setting a single thread resolved two of the six flaky tests (4\% of all reports) related to multi-threading. For example, in pull request \#780 of \texttt{qiskit-experiments}, developers observe occasional timeout issues when running multiple tests because multi-threading is disabled in certain environments. Though the flakiness may be resolved by disabling multi-threading, performance overheads may arise. Another example can be seen in Listing~\ref{lst:fix-thread} (pull request \#6539 in \texttt{qiskit-terra}), which shows the fix for the flaky test in Listing~\ref{lst:multi-thread} by disabling the parallelization. 

\begin{lstlisting}[language=Python, label={lst:fix-thread},
caption=The fix for the flaky test in Listing~\ref{lst:multi-thread}.]
- sphinx-build -W -b html -j auto docs/ docs/_build/html {posargs}
+ sphinx-build -W -b html docs/ docs/_build/html {posargs}
\end{lstlisting}

\subsubsection{Adjust Tolerance}\label{sec:increase-tolerance}

For flaky tests related to floating point operations, one can adjust tolerance. Three out of five reports, 6\% of all the reports, are fixed this way. For example, one can increase the tolerance manually, e.g., by doubling the tolerance value as shown in Listing~\ref{lst:fix-double}.  However, hard-coded loose tolerances may result in increased false negative test results, affecting the correctness of the program. 

\begin{lstlisting}[language=Python, label={lst:fix-double}, caption=An example of tolerance increase from pull request \#8820 of \texttt{qiskit-terra}.]
- np.testing.assert_allclose(result.values, [-1.307397243478641], rtol=0.05)
+ np.testing.assert_allclose(result.values, [-1.307397243478641], rtol=0.1)
\end{lstlisting}

Setting dynamic tolerance may be a more robust solution. For example, to fix the issue shown in Listing~\ref{lst:netkit_assert}, the developers compute the tolerance level dynamically, as shown in Lines 2 and 3 of Listing~\ref{lst:fix_netkit_assert}.

\begin{lstlisting}[language=Python, label={lst:fix_netkit_assert}, caption=An example of tolerance increase for Listing~\ref{lst:netkit_assert}.]
- assert driver.energy.mean == approx(ma.expect(ha).mean, abs=1e-5)
+ tol = driver.energy.error_of_mean * 5
+ assert driver.energy.mean == approx(ma.expect(ha).mean, abs=tol)
\end{lstlisting}

Developers also round the actual value to combat numeric errors (e.g., pull request \#4835 of \texttt{qiskit-terra}), or remove a flaky test case altogether if it is deemed non-essential (e.g., pull request \#8582 of \texttt{qiskit-terra}).

\subsubsection{Add Exception Handler}\label{sec:add_exception_handle}
Flaky tests caused by unhandled exceptions can be mitigated by adding an exception handler or removing the exception (all three reports, or 6\% of all the reports, are fixed this way). For example, developers add an \texttt{if} condition to remove any unexpected negative coefficients in pull request \#399 of \texttt{Microsoft/QuantumLibraries} (see Listing~\ref{lst:fix-exception}). 

\begin{lstlisting}[language=Python, label={lst:fix-exception},
caption=The fix for the flaky test in Listing~\ref{lst:unhandled}.]
- ApplyXorInPlace(keepCoeff[idx], keepCoeffRegister);
+ if (keepCoeff[idx] >= 0) {
+   ApplyXorInPlace(keepCoeff[idx], keepCoeffRegister);
+ }
\end{lstlisting}

\subsubsection{Synchronize}\label{sec:synchronize}
We observe one (i.e., 2\% of all the reports) network-related flaky test report, i.e., issue \#584 and pull request \#588 in \texttt{qiskit-ibm-runtime}. Here, tests for callback functions are dependent on socket tests. However, sometimes those callback tests finish before the socket tests, which causes flakiness. Developers increase callback test iterations so that socket tests have sufficient time to finish first (see Listing~\ref{lst:fix-network}). This may be a suboptimal solution as the number of iterations is hard-coded. The ideal solution would be to synchronize callback tests with the completion of socket tests.

\begin{lstlisting}[language=Python, label={lst:fix-network},
caption=The fix for the flaky test in Listing~\ref{lst:network}.]
- job = self._run_program(service, iterations=1, callback=result_callback)
+ job = self._run_program(service, iterations=10, callback=result_callback)
\end{lstlisting}

\subsubsection{Use Keys for Order}\label{sec:use_keys_order}
For flaky tests caused by unordered Python dictionaries, developers can use key values for ordering instead of using the insertion order. 
In Python 3.8 and later, dictionaries are order-preserved, so upgrading the Python environment can solve the problem. However, Python upgrades can cause dependency issues. One flaky test report (i.e., 2\% of all reports) had this cause and fix.

\section{Threats to Validity}\label{sec:threats}

Validity threats are classified according to~\cite{wohlin2012experimentation,yin2009case}.

\textbf{Internal and construct validity.} Data harvesting and cleaning are error-prone processes. The data were collected manually. Several co-authors independently examined the search results of flaky tests, then jointly reviewed and discussed the findings to finalize the list. The 10 keywords used to select the issue reports (discussed in Section~\ref{sec:empirical}) can have false negatives. Therefore, we underestimate the number of reports related to flaky tests. Yet, even this limited set of keywords proves that flaky tests exist in QC programs. In the future, we will extend this research to automated methods and tools to determine whether the reports are related to flakiness. 

\textbf{External and conclusion validity.}
Generally, software engineering studies suffer from real-world variability, and the generalization problem can only be solved partially~\cite{wieringa2015six}. We need to generalize a theoretical population and understand its architectural similarity relation to build a theory~\cite{wieringa2015six}. Although 14 open-source projects are used in this study, our findings may not generalize to other projects. However, based on our findings in this pilot study, the same empirical examination can be conducted on other quantum software products using well-designed and controlled experiments. Through such future research, we hope the community will expand our taxonomy of causes and identify patterns that will eventually lead to a general theory of flaky tests in quantum computing.

\section{Related Work}\label{sec:related}
The literature on testing and debugging quantum programs is growing. A quantum program is challenging to test because of quantum mechanics~\cite{miranskyy2019testing}. Software engineering principles are being applied to quantum program testing and debugging~\cite{miranskyy2019testing, miransky2020bug, miranskyy2021testing}; see~\cite{zhao2020quantum,DESTEFANO2022111326} for a comprehensive overview of quantum software engineering research work.

The community takes multiple approaches to tackle the challenge. Testing QC programs may be simplified by adding assertion checks to the code~\cite{huang2019statistical,li2020projection,liu2020quantum,ali2021assessing} or, in some cases, introducing debugging tricks, such as extracting classical information~\cite{miransky2020bug}. The identification of bug patterns in quantum programs can assist in defect analysis and categorization~\cite{zhao2021identifying}. We can also adapt classical fuzzy testing techniques~\cite{wang2018quanfuzz} or perform property-based testing~\cite{honarvar2020property}.
Additionally, quantum programs can be debugged in simulators on a CC (frameworks, such as Qiskit and Q\# readily provide such an option), but they can be used only for small problems~\cite{miranskyy2019testing,Usaola20}.

As far as we know, there has been no study of flaky tests in quantum programs. As discussed in Section~\ref{sec:cause}, Luo et al.~\cite{luo2014empirical} summarize 10 common causes of flakiness in software for CCs; four of them overlap with ours. Similar taxonomies for CC software have been presented by~\cite{parry2021survey,gruber2021empirical, dutta2020detecting}, which are complementary to ours.

\section{Conclusions and Future Work}\label{sec:conclusions}
This paper examines flakiness in 14 quantum programs. We detect 46 flaky test reports in 12 of these programs, which means at least 0.26\% to 1.85\% of issue reports in those programs are related to flakiness. Additionally, we identify eight groups of causes of flaky tests in QC code and seven common fixes. The final observation is that quantum programmers use only some recent software engineering techniques developed to deal with flaky tests. We hope that these findings will assist researchers and developers in mitigating the risk of flakiness when designing and testing quantum programs.

There are several ways to expand this work. In the future, we will explore additional repositories and keywords to identify more flaky test patterns in QC. In addition, we plan to compute the fraction of test cases affected by flakiness to perform direct statistical comparisons between CC and QC programs.

Our long-term objectives are 1) to automate the manual approaches to detect flaky tests in quantum programs based on the bug patterns identified in this paper and 2) to develop methods and tools to automatically fix QC flaky tests based on the fix patterns we have identified. For the first objective, we will develop an automated tool for each quantum flaky test category and integrate them into a framework that covers the most popular flaky test categories. As for the second objective, we will follow a similar path as the first one, i.e., we will develop individual automated fix methods and tools followed by a more comprehensive framework to fix flaky tests. These two objectives will combine organically to provide an efficient solution to mitigate flakiness in quantum programs.

We hope the community can use our findings to improve methods for detecting flakiness in the text of bug reports, automatically identifying flaky tests, suggesting mitigation strategies, and enriching flaky test datasets. Moreover, by raising awareness of the similarities and differences between classical and quantum programs, we would like to encourage quantum programmers to adopt some of the tools developed by the software engineering community to improve quantum software quality.

\bibliography{references} 
\bibliographystyle{IEEEtran}

\end{document}